\documentclass[12pt]{iopart}
\usepackage{iopams}
\usepackage{graphicx}
\usepackage{epstopdf}
\newcommand{\mytilde}{\raise.17ex\hbox{$\scriptstyle\mathtt{\sim}$}} 

\begin{document}

\title[UV LED charge control of an electrically isolated proof mass in a GRS at 255 nm]{UV LED charge control of an electrically isolated proof mass in a Gravitational Reference Sensor configuration at 255 nm}

\author{K~Balakrishnan$^1$,
  K-X~Sun$^2$,
  A~Alfauwaz$^4$,
  A~Aljadaan$^4$,
  M~Almajeed$^4$,
  M~Alrufaydah$^4$,
  S~Althubiti$^4$,
  H~Aljabreen$^4$,
  S~Buchman$^1$,
  R~L~Byer$^1$,
  J~Conklin$^1$,
  D~B~DeBra$^1$,
  J~Hanson$^5$,
  E~Hultgren$^1$,
  T~A~Saud$^4$,
  S~Shimizu$^1$,
  M~Soulage$^3$, and
  A~Zoellner$^1$
}

\address{$^1$ W.W. Hansen Experimental Physics Lab,
  Stanford University, Stanford, CA 94305}
\address{$^2$ University of Nevada at Las Vegas, Las Vegas, NV 89154}
\address{$^3$ NASA Ames Research Center, Moffett Field, CA 94035}
\address{$^4$ King Abdulaziz City for Science and Technology, Riyadh,
  Saudi Arabia 11442}
\address{$^5$ CrossTrac Engineering, Inc. Sunnyvale, CA 94089}

\ead{karthikb@stanford.edu}

\begin{abstract}
Precise control over the potential of an electrically isolated proof mass is necessary for the operation of devices such as a Gravitational Reference Sensor (GRS) and satellite missions such as LISA.  We show that AlGaN UV LEDs operating at 255 nm are an effective substitute for Mercury vapor lamps used in previous missions because of their ability to withstand space qualification levels of vibration and thermal cycling.  After 27 thermal and thermal vacuum cycles and 9 minutes of 14.07 g RMS vibration, there is less than 3\% change in current draw, less than 15\% change in optical power, and no change in spectral peak or FWHM (full width at half maximum).  We also demonstrate UV LED stimulated photoemission from a wide variety of thin film carbide proof mass coating candidates (SiC, Mo$_{2}$C, TaC, TiC, ZrC) that were applied using electron beam evaporation on an Aluminum 6061-T6 substrate.  All tested carbide films have measured quantum efficiencies of 3.8-6.8$\times10^{-7}$ and reflectivities of 0.11-0.15, which compare favorably with the properties of previously used gold films.  We demonstrate the ability to control proof mass potential on an 89 mm diameter spherical proof mass over a 20 mm gap in a GRS-like configuration.  Proof mass potential was measured via a non-contact DC probe, which would allow control without introducing dynamic forcing of the spacecraft.  Finally we provide a look ahead to an upcoming technology demonstration mission of UV LEDs and future applications toward charge control of electrically isolated proof masses.
\end{abstract}


\section{Introduction}
A Gravitational Reference Sensor (GRS) (\cite{Sun2006},\cite{sun2009},\cite{sun2011}) has been developed for space-based missions that require picometer or higher precision proof mass position measurement.  The GRS is a sensor for drag free control (\cite{lange1964thesis},\cite{triad1974},\cite{debra1999},\cite{debra2011}) that uses optical sensing techniques to determine the spacecraft position relative to the proof mass.  This allows the proof mass to float with near zero stiffness~\cite{Sun2006}.  The spacecraft housing blocks many disturbances including atmospheric drag, solar wind, and mitigates thermal effects, (\cite{higuchi2009},\cite{alfauwaz2011}) which otherwise may lead to the proof mass path deviating from the geodesic.  However, highly energetic particles are still capable of penetrating through the spacecraft outer surface and charging the proof mass, either directly or via secondary electron emission (\cite{Burke1980},\cite{Thomson2003},\cite{Miyake2006}).  This in turn leads to proof mass charging rates on the order of \mytilde+50 e$^{-}$/s~\cite{Sumner2009} depending on spacecraft size, shielding provided by the proof mass housing, and orbit.  The caging and uncaging process can also leave behind residual charges on the proof mass.  Such charging leads to an electrostatic disturbance force that would corrupt the signal necessary for both the scientific measurement and drag-free control.

\begin{figure}[htb]
	\centerline{\includegraphics{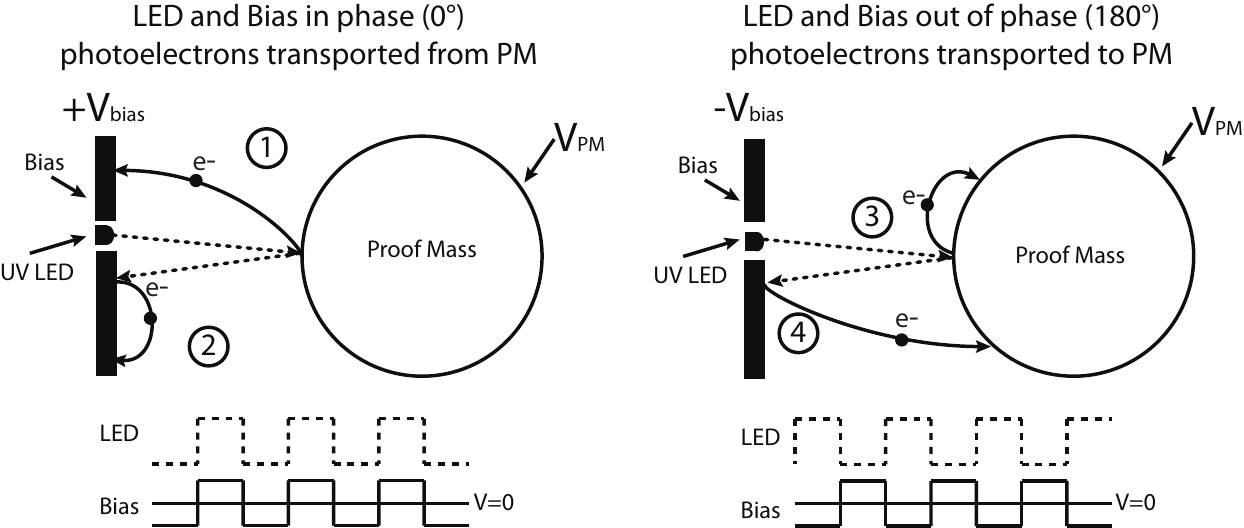}}
	\caption{Schematic showing the basic operation of AC charge management. The solid lines describe the electron path while the dashed lines describe the UV path. When the bias plate voltage relative to the housing ($V_{bias}$) is positive, (1) photoelectrons generated from the proof mass and (2) photoelectrons generated from the bias plate travel to the bias plate leading to an increase in the proof mass potential ($V_{PM}$).  When $V_{bias}$ is negative, (3) photoelectrons generated from the proof mass and (4) photoelectrons generated from the bias plate travel to the proof mass leading to a decrease in $V_{PM}$.}
	\label{fig:ACCMSbasics}
\end{figure}

Charge management is achieved through photoemission; missions such as Gravity Probe B~\cite{buchman1995} and LISA Pathfinder~\cite{Sumner2009} have used the 254 nm UV line of mercury lamps as the light source.  Deep UV LEDs operating at 255 nm have been identified as a new method to mitigate proof mass charging~\cite{Sun2006a}.  Compared to Hg lamps, UV LEDs are smaller and lighter, consume less power, have a wider spectrum selection, and a much higher dynamic range, with at least an order of magnitude improvement in each performance area.  The power output is also very stable, with a lifetime $>$ 30000 hours.  By using UV LEDs for AC charge transfer, charge management can be performed outside the science band.  We have demonstrated an AC charge management system in a GRS-like configuration using a single bias plate with large gap size, UV LED light source, and a single large spherical proof mass as shown in Figure~\ref{fig:ACCMSbasics}.

AC charge management can be split into two cases: in phase (positive charge transfer) or out of phase (negative charge transfer).  In both cases, UV light is directed at the proof mass and photoelectrons are generated from its surface.  Some of the UV light reflects back to the bias plate and photoelectrons are generated from the bias plate surface. In the positive transfer case, both drive signals are in phase so $V_{bias}$ is positive while the LED is turned on; generated photoelectrons are pulled towards the bias plate.  When $V_{bias}$ is negative, the LED is off and no photoelectrons are generated.  In the negative transfer case, because the drive signals are out of phase, $V_{bias}$ is negative when photoelectrons are generated.  Thus, electrons are pushed from the bias towards the proof mass.  When $V_{bias}$ is positive, the LED is off and no photoelectrons are generated.  The rate of charging depends on UV optical power, coating properties such as workfunction, quantum efficiency, and reflectivity, and the surface roughness of the proof mass and bias plate.

\section{UV source properties and testing}

\begin{figure}[htb]
	\centerline{\includegraphics{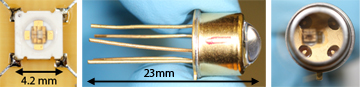}}
	\caption{UV LEDs in various packaging styles.  From left to right: Surface Mount (SMD) affixed on a printed circuit board, Hemispherical (HS), and Tall Flat Window (TFW).}
	\label{fig:LEDphotos}
\end{figure}

The UV LED source considered for GRS charge management is an AlGaN based device supplied by Sensor Electronic Technology, Inc.  The UV LEDs are available in several packages, such as those shown in Figure~\ref{fig:LEDphotos}, allowing them to be easily integrated inside a GRS housing. They contain both an LED source as well as a witness photodiode, allowing for real-time monitoring of the UV output.

We have performed extensive tests including lifetime, radiation, and MIL-STD-1540E~\cite{Perl2006} level thermal and vibration to validate device robustness for spaceflight.  Lifetime tests in vacuum (currently $>$30000 hours) and nitrogen ($>$12000 hours), high fluence proton irradiation (63 MeV at a fluence of 2$\times10^{12}$ protons/cm$^{2}$)~\cite{Sun2009a}, thermal vacuum (30 cycles at -34$^{\circ}$C to +71$^{\circ}$C), and vibration (14.07 g RMS for 3 minutes per axis)~\cite{balakrishnan2011} have demonstrated the robustness of UV LEDs.

\begin{figure}[htb]
	\centerline{\includegraphics{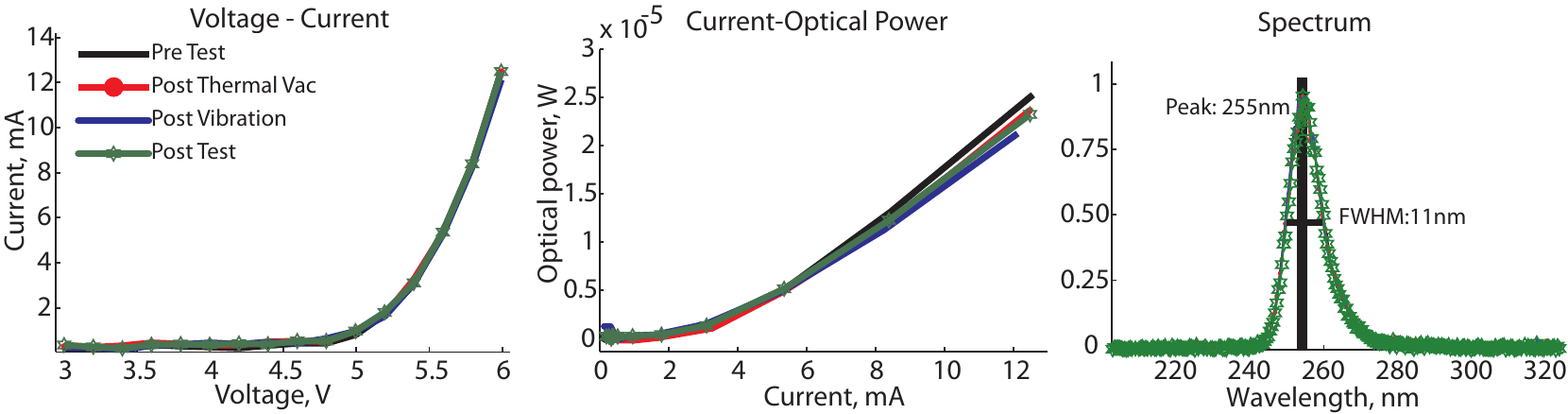}}
	\caption{Characteristic performance plots of an uncollimated UV LED taken during MIL-1540 level laboratory testing.  Figures show data taking before testing, after thermal vacuum cycling, after shake, and after thermal cycling (post test). From left: Voltage (V) vs. Current (mA), Current (mA) vs. Optical Power ($\mu$W), and Spectrum}
	\label{fig:LEDprops}
\end{figure}

Representative results from the MIL-1540 laboratory testing are shown in Figure~\ref{fig:LEDprops}.  The V-I curve was generated using an Agilent E3631A power supply, optical power was measured with a Newport 1931-C power meter with a 918D-UV3 detector head, and spectrum was measured with an OceanOptics MayaPro spectrometer.

The V-I (voltage-current) curve for a diode is the fundamental measure of its PN junction characteristics and a family of well behaved V-I curves is a good indication of diode chipset quality.  During MIL-1540 level testing, the semiconductor chipset performance remains stable through all thermal and vibration tests, indicating that the electrical properties of the chipset are effectively unchanged.  In the 5-10 mA range where most charge management will be performed, there is an increase current draw of less than 3\% in the tested diodes.  Similarly, there is no shift in spectral peak or FWHM (full width at half maximum) .  This level of robustness indicates that UV LEDs are an ideal candidate for spacecraft charge mitigation in applications currently using mercury lamps.

\section{Proof Mass Coatings}
\subsection{Coating requirements and candidate selection}
Selection of a proper proof mass coating is critical for both charge management and general performance of the GRS.  The photon energy $E_{photon}$ from a 255 nm UV LED source is 4.86 eV, and this sets the upper bound on the candidate coating workfunctions,~$\phi$.  The selected coating should also act as a protective layer, allowing the proof mass to retain surface optical and geometric properties during caging, uncaging, thermal cycling, and during any possible collision with GRS housing walls.  The coating must also strongly adhere to the proof mass during launch vibrations of up to 50 g and during thermal cycling; there must also be minimal adhesion between the surface coating and the GRS housing.  In addition, the coating must be highly reflective in the IR (1064 and 1550 nm) for optical proof mass readout and be electrically conductive to help minimize patches.

Several materials were selected for evaluation based on the above criteria.  Au was chosen as a default coating based on its widespread use in spacecraft~\cite{Shaul2008} while Nb was chosen because of flight heritage as coating for the GP-B gyroscopes~\cite{buchman1995}. In addition, several carbides were selected because of their mechanical toughness \cite{Storms1964} and workfunction (\cite{III/17A-22A-41A1b},\cite{Lide},\cite{Oshima1981},\cite{Wilson1967},\cite{Gotoh2003}).  Carbides are commonly used in applications such machine tool coatings and disk brakes where mechanical toughness is a concern, making them an attractive choice for meeting the proof mass protection requirement.  The full list of candidate coatings is: Au, Nb, Mo$_{2}$C, SiC, TaC, TiC, and ZrC.

\subsection{Sample preparation}
The substrates used for coating characterization were 1.5 in. (38.1 mm) Al 6061-T6 squares with a thickness of 0.125 in. (3.175 mm).  The plates were machined to a surface roughness of Ra64 (1.62 $\mu$m) then cleaned via an HF etch to remove surface impurities.

The carbide materials were obtained in pellet form with diameters ranging from 2 to 4 mm, then coated via E-beam evaporation onto the Al plates.  Au and Nb materials were coated using round evaporation targets with 1.25 in. (31.75 mm) diameter.  The final thickness of the Nb and carbide films was 1000 \AA (100 nm).  The total film thickness of the Au sample was 1150 \AA (115 nm): 1000 \AA\ (100 nm) Au on a 150 \AA\ (15 nm) Ti sticking layer.  

\begin{figure}[htb]
	\centerline{\resizebox{!}{2.5in}{\includegraphics{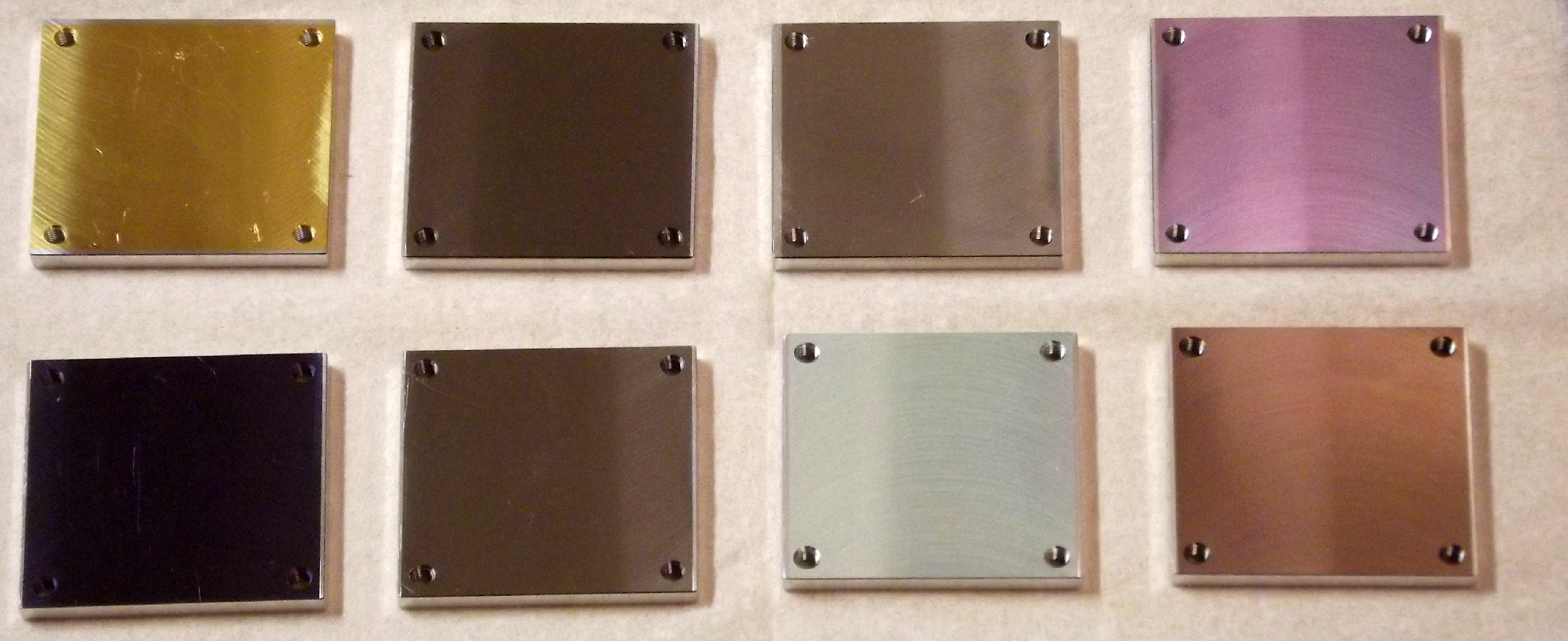}}}
	\caption{Coated aluminum samples. Top row (from left): Au, Nb, Ir, SiC. Bottom row (from left): TiC, Mo$_{2}$C, ZrC, TaC. A color version of the image is available in the electronic version of this document.}
	\label{fig:coatingSamples}
\end{figure}

\subsection{Measurement of coating quantum efficiency}
In previous mesurements~\cite{buchman1995}, UV light is directed through a mesh onto the sample surface. By biasing the sample negative and the mesh positive, photoelectrons are captured by the mesh.  The photocurrent is then equal to the current necessary to hold the mesh at a fixed potential.  However, because the incident light from the UV LED is randomly polarized, photoelectrons are emitted in random directions.  If a mesh is used, some photoelectrons may be left uncaptured even with the voltage bias.

\begin{figure}[htb]
	\centerline{\includegraphics{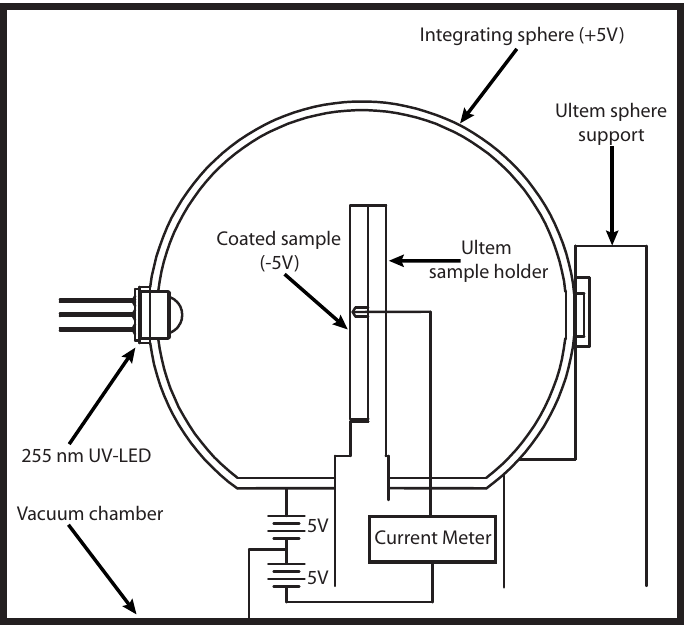}}
	\caption{Schematic of experimental setup used to measure thin film coating UV quantum efficiency.}
	\label{fig:QEmeasSetup}
\end{figure}

In an alternative approach shown in Figure~\ref{fig:QEmeasSetup}, the sample is placed inside of a hollow integrating sphere.  UV light from an LED at 255 nm (FWHM of 11 nm) is directed toward the coated sample through an opening in the sphere, with an incident UV power of 50 $\mu$W.  The coated sample is biased to -5 V and the surrounding sphere is biased to +5 V.  Because the sphere surrounds the sample, most photoelectrons are captured.  The resulting current is measured using a Keithley 6485 picoammeter.  To limit leakage currents, both the sphere and sample are isolated from ground via high aspect ratio Ultem-1000 posts with bulk resistivity $\rho=10^{17}$ $\Omega$-cm.  The measurement is performed in vacuum below $10^{-4}$ Torr, achieved using dry pumping equipment to prevent sample contamination.  Quantum efficiency is computed via:

\begin{equation}
QE = \frac{N_{e}}{N_{\nu}}
\end{equation}

\noindent where $N_{\nu}$ is the number of photons absorbed by the thin film coating and $N_{e}$ is the number of electrons emitted.

\subsection{Coating reflectivity}

In order to decrease the proof mass potential, electrons must be transported to the proof mass from the housing.  This can be done in several ways, including direct illumnation of the housing, an electron gun, or by using reflected UV light.  The third option is used in the GRS design.  A UV LED directly illuminates the proof mass, which fully intercepts the beam.  Some light reflects back to the housing and generates the electrons necessary for negative charge transfer.  For this design to work, the coating must have a reflectivity at 255 nm that is sufficiently high to generate photoelectrons from the housing walls.

Measurement of the coating reflectivity was performed by shining light from a single collimated hemispherical lens UV LED at a coated sample held rigidly on an optical bench.  Reflected power was measured directly with a Newport 1931C power meter and 918D-UV3 detector head and calibrated using a coated Ir sample with known reflectivity of 0.6 at 255 nm.  Angle of incidence was fixed at 45$^{\circ}$.  Reflectivity was computed at 255~nm via:

\begin{equation}
\rho(\lambda) = \frac{P_{reflected}(\lambda)}{P_{incident}(\lambda)}
\end{equation}

\noindent where $P_{reflected}$ is the reflected optical power as measured by the power meter, and $P_{incident}$ is the incident optical power as derived through calibration with the Ir sample.

\subsection{Properties of considered coatings}
Both the measured and tabulated properties of the considered coatings are shown in Table~\ref{tab:coatprop} .  I$_{Sample}$ is the current required to hold the coated sample at a constant potential of -5 V relative to the vacuum chamber during the QE measurement.  The variables QE$_{mean}$ and $\sigma_{QE}$ are the QE mean and standard deviation as computed from 10 measurements; $\rho$ is reflectivity calculated at $\lambda$=255 nm, and $\phi$ is workfunction tabulated from several sources.

\begin{table}[hbpt]
  \centering
  \caption{Optical and electrical properties of various metal and carbide thin film coatings on an Al 6061-T6 substrate. The $\dag$ indicates properties tabulated from other sources; all other values were measured.}
    \begin{tabular}{cccccccc}
    Material & I$_{Sample}$(pA) & QE$_{mean} $ & $\sigma_{QE}$ & $\rho$(255 nm) & $\phi$ (eV)$^\dag$ \\
    \hline
    Au    			& 3.49 & 3.4E-07 & 3.5E-08 & 0.17 & 5.47~\cite{III/17A-22A-41A1b}		   \\
    Nb    		  & 5.78 & 5.6E-07 & 4.5E-08 & 0.17 & 4.30~\cite{Lide}							 \\
    SiC  		    & 4.46 & 4.3E-07 & 1.3E-08 & 0.12 & 4.80~\cite{III/17A-22A-41A1b}	 \\
    TiC   			& 4.60 & 4.5E-07 & 2.5E-08 & 0.15 & 3.80~\cite{Oshima1981}				 \\
    ZrC   			& 3.95 & 3.8E-07 & 7.8E-09 & 0.11 & 3.70~\cite{Wilson1967}				 \\
    Mo$_{2}$C   & 6.99 & 6.8E-07 & 1.5E-08 & 0.15 & 4.74~\cite{Wilson1967}         \\
    TaC   			& 6.51 & 6.3E-07 & 2.0E-08 & 0.13 & 5.0~\cite{Gotoh2003}           \\
    \end{tabular}%
  \label{tab:coatprop}%
\end{table}%

There is little variation between the quantum efficiencies and reflectivities of all coating types. Gold has the lowest QE of the tested materials, and Mo$_{2}$C has the highest; they only vary by a factor of two.  Similarly, all reflectivities lie between 0.11 and 0.17.  This indicates that all selected coatings will be suitable for charge management.  Using a standard pull test using commercially available Scotch tape, all films remained secured to the aluminum substrate.

In the future, quantitative tests of thin film hardness, surface resistivity, optical reflectivity over a range of $\lambda$, and surface adhesion to the substrate and other coatings will be performed.

\section{AC (Active) Charge Management Experiment}
\subsection{Experimental setup}
\label{sec:cmsSetup}

\begin{figure}[hbtp]
	\centerline{\includegraphics{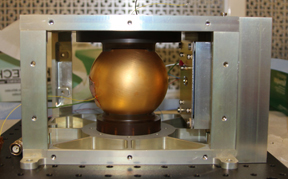}}
	\caption{Prototype research system showing single bias plate, coated sphere, and Ultem holding tubes containing floating probes.  The wire attached to the proof mass was used during early tests to calibrate the PM potential with the potential measured by the floating probe, but was removed during actual experimental runs.}
	\label{fig:protoSyst}
\end{figure}

\begin{figure}[htb]
	\centerline{\includegraphics{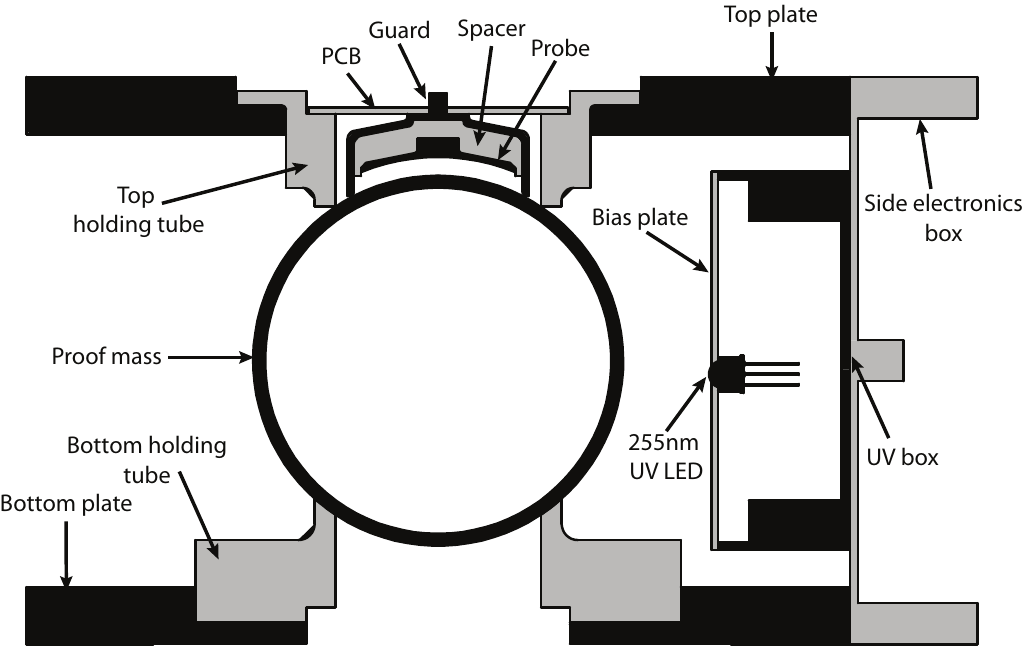}}
	\caption{Cutaway schematic of AC charge management experiment structure. Total proof mass capacitance to ground is 17 pF.}
	\label{fig:rmSchematic}
\end{figure}

\begin{figure}[htb]
	\centerline{\includegraphics{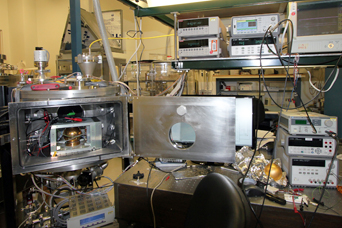}}
	\caption{Prototype research system inside vacuum chamber.}
	\label{fig:protoLab}
\end{figure}

The lab demonstration was performed using commercially available test equipment.  A single 255 nm UV LED was used as the UV source.  A dual channel function generator (Agilent 33522A) modulated both the LED and Bias allowing precise control of relative phase.  Proof mass potential was read back using an electrometer (Keithley 6514) in voltage mode using guarded inputs. The integrated photodiode response was measured with a picoammeter (Keithley 6485).  A photograph showing the experiment as well as driving electronics in the laboratory is shown in Figure~\ref{fig:protoLab}.

A schematic showing the experimental setup is shown in Figure~\ref{fig:rmSchematic}, and a photograph of the experimental structure is shown in Figure~\ref{fig:protoSyst}.  The proof mass is a single 3.5 in. (88.9 mm) diameter hollow Aluminum 6061-T6 sphere with 0.125 in. (3.175 mm) wall thickness.  The exterior of the sphere was cleaned via an HF etch.  A 200 \AA (20 nm) Ti sticking layer was coated over the entire surface followed by a 1500 \AA (150 nm) Au layer, both via E-beam evaporation.  The Ti improves Au adhesion and prevents alloying between the Al and Au layers. Support is provided by two Ultem-1000 holding tubes.

Bias is provided via a single Aluminum 6061-T6 square plate measuring 92 mm on an edge, located 20 mm away from the closest point of the proof mass.  The bias plate is unpolished (standard extruded roughness of 3.5-7.5 $\mu$m)~\cite{mcmaster} and coated via E-beam evaporation with a 200 \AA\ (20 nm) Ti sticking layer and 1500 \AA\ (150 nm) Au photoemission layer.

Proof mass potential is measured using a single 38 mm diameter gold coated aluminum probe disk housed inside a guard shell.  The probe is curved so that its surface is always parallel to the proof mass.  As a result, the magnitude of the electric field caused by the proof mass is equal along most of the probe.  This guard is driven to the probe potential and helps shield the measurement from intereference by driving electronics.  The probe is located 4 mm above the proof mass and gives a direct DC measurement of the electric field immediately surrounding the proof mass.

\subsection{Results}
\begin{figure}[htb]
	\centerline{\includegraphics{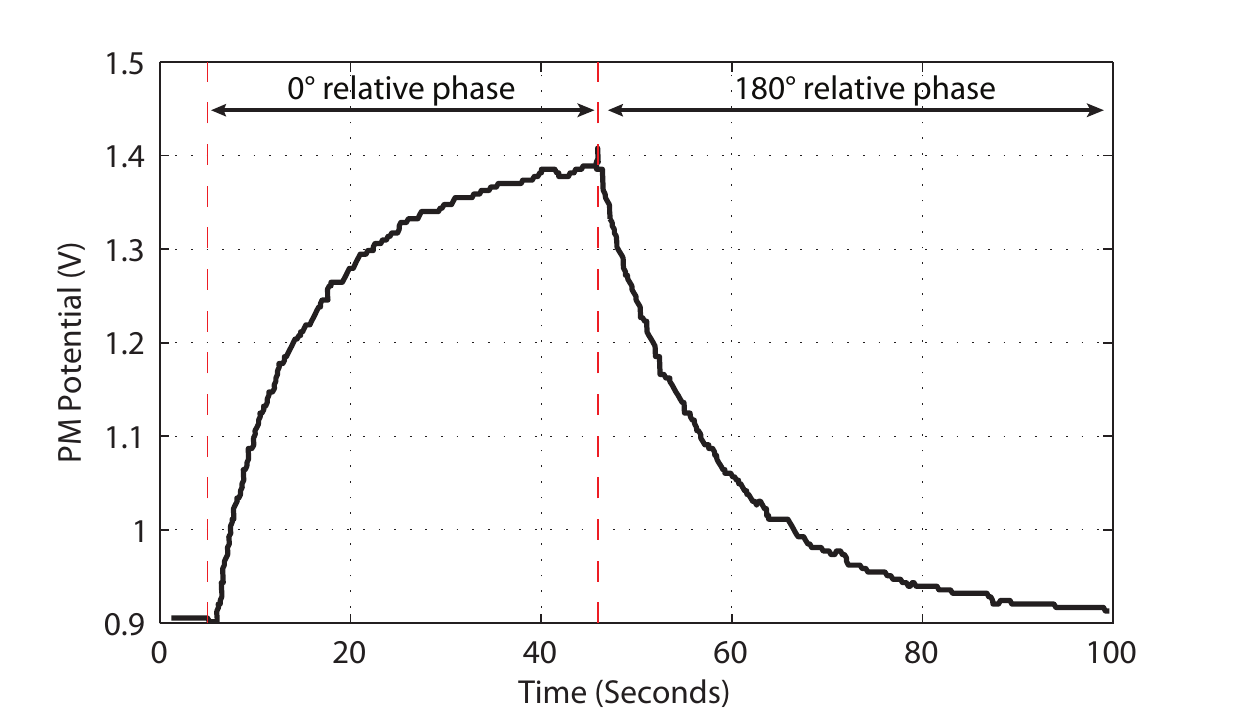}}
	\caption{Proof mass potential measured using a floating probe in the AC Charge Control regime.  Potential is measured relative to the system housing which is held at ground.  Incident UV power is 10 $\mu$W modulated at 100~Hz, 50\% duty cycle, with a 3.0 V$_{pp}$ bias. Proof mass capacitance to ground is 17 pF.}
	\label{fig:singleFlipACCms}
\end{figure}

During the experimental run shown in Figure~\ref{fig:singleFlipACCms}, a single LED was modulated at 100~Hz with a 50\% duty cycle and driven at 6.4 mA, resulting in an optical power output of 10 $\mu$W. The bias was also modulated at 100~Hz, 3.0 V$_{pp}$.  Both LED and bias were initially in phase with each other (0$^\circ$) leading to positive charge transfer; upon reaching the space-charge limited regime, the bias phase was flipped to 180$^\circ$ leading to negative charge transfer.  Note that positive charging occurs more quickly (40 seconds) than negative charging (55 seconds).  The rate at which the proof mass potential changes can be reduced by decreasing the LED optical power or decreasing the duty cycle during which the LED is switched on.  Previous tests have shown that the UV LED is capable of performing charge management when being driven at 10 kHz~\cite{Sun2006a} which means that the duty cycle can be reduced from 50\% to at least as low as 0.5\% with no impact on LED performance capabilities.

\section{Applications and Flight}
\subsection{Updated charge management structure design}
\begin{figure}[htb]
	\centerline{\includegraphics{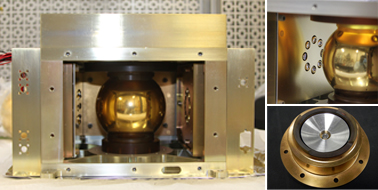}}
	\caption{Research model of UV LED payload developed at Stanford.  Clockwise from left: view of whole research structure, 8 UV LEDs directed towards proof mass through openings in bias plate, charge amp simulator and contact probe installed in gold coated Ultem holding tube.}
	\label{fig:researchModel}
\end{figure}

Several modifications have been made at Stanford in an updated prototype, shown in Figure~\ref{fig:researchModel}:

\begin{itemize}
	\item The number of LEDs has been increased from 1 to 16, allowing for increased redundancy.  This increase also allows a much higher optical power to be directed onto the proof mass, increasing the possible charging and discharging rate.
	\item Both surface mount and flat window style LEDs are present.
	\item The floating probe has been replaced with a contact probe and charge amplifier with an input impedence of 10$^{14}$ $\Omega$.
	\item The number of bias plates has been increased from 1 to 4. All bias plates are increased in size.
	\item The proof mass has been polished to a mirror finish before coating.
	\item Both Ultem holding tubes have been gold coated (with the exception of a small pad that contacts the proof mass) to provide additional electrical shielding for the contact probe.
\end{itemize}

\subsection{Flight demonstration of UV LED}

A spacecraft demonstration of UV LEDs and UV LED charge management based on research done at Stanford is being developed jointly by KACST, and NASA Ames Research Center, with an expected launch date of early to mid 2013.  Mission lifetime is expected to be at least one month, during which time the ability for the UV LED to mitigate actual space-based charging and the effects of radiation on the UV LED device performance will be studied.  The goal of the mission is to increase the UV LED device to TRL-9 and the charge management system to TRL-7. The relevant TRL levels, defined in~\cite{Engineering2011}, are:

\begin{itemize}
	\item TRL 7: Prototype near or at planned operational system. Represents a major step up from TRL 6 by requiring demonstration of an actual system prototype in an operational environment (e.g., in an air-craft, in a vehicle, or in space).	
	\item TRL 9: Actual application of the technology in its final form and under mission conditions, such as those encountered in operational test and evaluation (OT\&E). Examples include using the system under operational mission conditions.	
\end{itemize}

\subsection{Application to a GRS}

\begin{figure}[htb]
	\centerline{\includegraphics{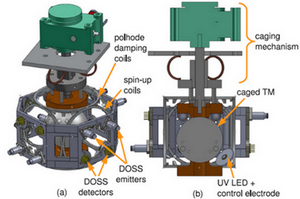}}
	\caption{CAD model of GRS concept showing key components.}
	\label{fig:MGRScad}
\end{figure}

The UV LED charge management system has a direct application in a Gravitational Reference Sensor (GRS) concept used for true drag-free flight, shown in Figure~\ref{fig:MGRScad}~\cite{Conklin2011}.  The proof mass is a single 70\%/30\% Au/Pt spinning 70 mm sphere coated in a carbide compound such as those described above.  A large 35 mm gap between proof mass and housing is chosen in order to improve acceleration noise performance.  During launch, the proof mass is held with a caging system designed to secure the mass through random vibration up to 14.07 g rms and shock up to 3000 g at payload separation.  After launch, the caging system is retracted and the proof mass is freed.  The proof mass is then spun up to $\approx$10 Hz using a rotating magnetic field.

The control signal for drag free control is provided via the Differential Optical Shadow Sensor (DOSS) which operates by using four parallel beam pairs for a redundant three-dimensional position measurement.  Thermal control of the proof mass is required to a level $< 10$ $\mu K$ and is provided using layers of highly conductive shields and and vacuum spacing~\cite{alfauwaz2011}. 

\section{Conclusions}

A LED source for charge management has been an area of great interest because of the advantages that it offers over traditional Hg lamp systems.  The research data show that the UV LED is capable of controlling proof mass potential and that this potential can be measured without introducing a disturbance force.  The increased variety of surface coatings provides a larger design space for the proof mass, allowing for better performance and higher survivability of the system.  We are interested in further research on coating properties, charge management techniques, and integration of the UV LED into a full GRS setup. This research will be applicable to missions requiring GRS-level precision such as LISA, as well as those requiring charge management of electrically isolated components.

\section*{Acknowledgements}
Karthik Balakrishnan was supported by the Department of Defense (DoD) through the National Defense Science \& Engineering Graduate Fellowship (NDSEG) Program.  The work described in this paper has been funded by the King Abdulaziz City for Science and Technology (KACST) and NASA Ames Research Center (ARC).  We would also like to thank:

\begin{itemize}
	\item Dale Gill and Tom Carver (Stanford) for providing advice on coating selection and characterization
	\item Timothy R. Palmer (Massachusetts Institute of Technology) for advice on coating application and substrate preparation
	\item Robert Ricks and Kuok Ling (NASA ARC) for low current measurement techniques
	\item Dave Muselman (DP Precision) for DFM suggestions on the AC charge management research structure 
\end{itemize}

\section*{References}
\bibliography{paper}

\begin{thebibliography}{10}

\bibitem{Sun2006}
Ke-Xun Sun, Graham Allen, Scott Williams, Saps Buchman, Dan DeBra, and Robert
  Byer.
\newblock {Modular Gravitational Reference Sensor: Simplified Architecture to
  future LISA and BBO}.
\newblock {\em Journal of Physics: Conference Series}, 32:137--146, 2006.

\bibitem{sun2009}
Ke-Xun Sun, Saps Buchman, Robert Byer, Dan DeBra, John Goebel, Graham Allen,
  John~W Conklin, Domenico Gerardi, Sei Higuchi, Nick Leindecker, Patrick Lu,
  Aaron Swank, Edgar Torres, and Martin Trittler.
\newblock {Modular gravitational reference sensor development}.
\newblock {\em Journal of Physics: Conference Series}, 154:012026, 2009.

\bibitem{sun2011}
K.-X. Sun, A.~Alfauwaz, M.~Alrufaydah, H.~Altwaijry, K.~Balakrishnan,
  S.~Buchman, R.~L. Byer, J.~W. Conklin, D.~B. DeBra, J.~Goebel, E.~Hultgren,
  and A.~Zoellner.
\newblock {Modular Gravitational Reference Sensor (MGRS) Technology
  Development}.
\newblock In {\em Proceedings of the 8th International LISA Symposium}, Journal
  of Physics Conference Series, 2011.

\bibitem{lange1964thesis}
B.~Lange.
\newblock {\em {The Control and use of Drag-free Satellites}}.
\newblock {PhD thesis}, Stanford University, 1964.

\bibitem{triad1974}
{Staff Of The Space Department}, {Staff Of The Guidance}, and {Control
  Laboratory}.
\newblock {A Satellite Freed of all but Gravitational Forces: ''TRIAD I''}.
\newblock {\em Journal of Spacecraft and Rockets}, 11:637, September 1974.

\bibitem{debra1999}
D.~B. {DeBra}.
\newblock {Design considerations for drag free satellites}.
\newblock In {W.~M.~Folkner}, editor, {\em Laser Interferometer Space Antenna,
  Second International LISA Symposium on the Detection and Observation of
  Gravitational Waves in Space}, volume 456 of {\em American Institute of
  Physics Conference Series}, pages 199--206, December 1998.

\bibitem{debra2011}
D.~B. {DeBra} and J.~W. {Conklin}.
\newblock {Measurement of drag and its cancellation}.
\newblock {\em Classical and Quantum Gravity}, 28(9):094015, May 2011.

\bibitem{higuchi2009}
S.~{Higuchi}, K.-X. {Sun}, D.~B. {DeBra}, S.~{Buchman}, and R.~L. {Byer}.
\newblock {Design of a highly stable and uniform thermal test facility for MGRS
  development}.
\newblock {\em Journal of Physics Conference Series}, 154(1):012037, March
  2009.

\bibitem{alfauwaz2011}
A.~Alfauwaz and K.-X. Sun.
\newblock {Design and Modeling of Highly Stable and Uniform Thermal Enclosure
  for Precision Space Experiment}.
\newblock In {\em Journal of Physics Conference Series}, Proceedings of the 8th
  International LISA Symposium, 2011.

\bibitem{Burke1980}
E.~A. Burke.
\newblock {Secondary Emission from Polymers}.
\newblock {\em IEEE Transactions on Nuclear Science}, 27(6):1759--1764, 1980.

\bibitem{Thomson2003}
C.~Thomson, V.~Zavyalov, J.R. Dennison, and J~Corbridge.
\newblock {Electron emission properties of insulator materials pertinent to the
  International Space Station}.
\newblock In {\em 8th Spacecraft Charging Technology Conference}, Huntsville,
  AL, 2003.

\bibitem{Miyake2006}
Hiroaki Miyake, Kumi Nitta, Shinichiro Michizono, and Yoshio Saito.
\newblock {Secondary Electron Emission Measurement of Insulating Materials for
  Spacecraft}.
\newblock In {\em 2006 International Symposium on Discharges and Electrical
  Insulation in Vacuum}, pages 770--773. IEEE, 2006.

\bibitem{Sumner2009}
T~J Sumner, DNA Shaul, M~O Schulte, S~Waschke, D~Hollington, and H~Ara\'{u}jo.
\newblock {LISA and LISA Pathfinder charging}.
\newblock {\em Classical and Quantum Gravity}, 26(9):094006, May 2009.

\bibitem{buchman1995}
S.~{Buchman}, T.~{Quinn}, G.~M. {Keiser}, D.~{Gill}, and T.~J. {Sumner}.
\newblock {Charge measurement and control for the Gravity Probe B gyroscopes}.
\newblock {\em Review of Scientific Instruments}, 66:120--129, January 1995.

\bibitem{Sun2006a}
Ke-Xun Sun, Brett Allard, Saps Buchman, Scott Williams, and Robert~L Byer.
\newblock {LED deep UV source for charge management of gravitational reference
  sensors}.
\newblock {\em Classical and Quantum Gravity}, 23(8):S141--S150, 2006.

\bibitem{Perl2006}
E~Perl.
\newblock {Test requirements for launch, upper stage, and space vehicles}.
\newblock Technical report, The Aerospace Corporation, September 2006.

\bibitem{Sun2009a}
Ke-Xun Sun, Nick Leindecker, Sei Higuchi, John Goebel, Sasha Buchman, and
  Robert~L Byer.
\newblock {UV LED operation lifetime and radiation hardness qualification for
  space flights}.
\newblock {\em Journal of Physics: Conference Series}, 154:012028, 2009.

\bibitem{balakrishnan2011}
K.~Balakrishnan, E.~Hultgren, J.~Goebel, and K.-X. Sun.
\newblock {Space Qualification Test Results of Deep UV LEDs for AC Charge
  Management}.
\newblock In {\em 11th Spacecraft Charging Technology Conference}, poster
  presentation, September 2011.

\bibitem{Shaul2008}
D~N~A Shaul, H~M A R~A Ujo, G~K Rochester, M~Schulte, T~J Sumner, C~Trenkel,
  and P~Wass.
\newblock {Charge management for lisa and lisa pathfinder ´}.
\newblock {\em International Journal of Modern Physics D}, 17(7):993--1003,
  2008.

\bibitem{Storms1964}
E.K. Storms.
\newblock {A CRITICAL REVIEW OF REFRACTORIES}.
\newblock Technical report, Los Alamos Scientific Lab, Los Alamos, New Mexico,
  1964.

\bibitem{III/17A-22A-41A1b}
Authors III/17A-22A-41A1b and Editors of~the LB~Volumes.
\newblock {\em {Silicon carbide (SiC), work function}}.
\newblock SpringerMaterials - The Landolt-B\"{o}rnstein Database.

\bibitem{Lide}
David Lide, editor.
\newblock {\em {CRC Handbook of Chemistry and Physics}}.
\newblock CRC Press, 85 edition.

\bibitem{Oshima1981}
C.~Oshima, M.~Aono, T.~Tanaka, S.~Kawai, S.~Zaima, and Y.~Shibata.
\newblock {Clean TiC(001) surface and oxygen chemisorption studied by work
  function measurement, angle-resolved X-RAY photoelectron spectroscopy,
  ultraviolet photoelectron spectroscopy and ion scattering spectroscopy}.
\newblock {\em Surface Science}, 102(2-3):312--330, January 1981.

\bibitem{Wilson1967}
R.~G. Wilson.
\newblock {Vacuum Thermionic Work Functions and Thermal Stability of TaB2, ZrC,
  Mo2C, MoSi2, TaSi2, and WSi2}.
\newblock {\em Journal of Applied Physics}, 38(4):1716, 1967.

\bibitem{Gotoh2003}
Y.~Gotoh, H.~Tsuji, and J.~Ishikawa.
\newblock {Measurement of work function of transition metal nitride and carbide
  thin films}.
\newblock {\em Journal of Vacuum Science \& Technology B: Microelectronics and
  Nanometer Structures}, 21(4):1607, 2003.

\bibitem{mcmaster}
McMaster-Carr.
\newblock personal communication, 2012.

\bibitem{Engineering2011}
Assistant Secretary of Defense for~Research and Engineering.
\newblock {Technology Readiness Assessment ( TRA ) Guidance}.
\newblock Technical Report May, Department of Defense, 2011.

\bibitem{Conklin2011}
J~W Conklin, S~Buchman, V~Aguero, A~Alfauwaz, and Et. Al.
\newblock {LAGRANGE : LAser GRavitational-wave ANtenna at GEo-lunar Lagrange
  points (arXiv: 1111.5264)}.
\newblock 2011.

\end{thebibliography}

\end{document}